\documentstyle[11pt]{article}

\begin{document}

\newcommand{\beq}{\begin{equation}}
\newcommand{\eeq}{\end{equation}}
\newcommand{\sG}{i\sigma G}

\renewcommand{\Im}{\mbox{Im}\,}
\renewcommand{\Re}{\mbox{Re}\,}
\newcommand{\ra}{\rightarrow}
\newcommand{\La}{\overline{\Lambda}}
\newcommand{\Lam}{\Lambda_{QCD}}

\begin{titlepage}
\renewcommand{\thefootnote}{\fnsymbol{footnote}}

\begin{flushright}
\large{
TPI-MINN-94/17-T\\
UMN-TH-1253/94}
\end{flushright}
\vspace{0.4cm}
\begin{center} \LARGE
{\bf Theory of Pre-Asymptotic Effects in Weak Inclusive Decays }
\end{center}
\begin{center}
{\em Talk at the  Workshop ``Continuous Advances in
QCD"}\footnote{Extended version}
\end{center}
\begin{center}
{\em February 18 -- 20, 1994, TPI, Univ. of Minnesota, Minneapolis}
\vspace{0.4cm}
\\
{\bf M.A. Shifman}\\
{\normalsize  Theoretical Physics Institute, Univ. of Minnesota,
Minneapolis, MN 55455}
\\{\normalsize  e-mail address: SHIFMAN@VX.CIS.UMN.EDU}
\vspace{0.4cm}
\end{center}

\vspace{1cm}

\centerline{\Large\bf Abstract}

\vspace{.4cm}

I give an introduction to the theory of preasymptotic effects based
on the systematic OPE/HQET expansion in $1/m_Q$ where $m_Q$
is the heavy quark mass. The general idea is explained in two most
instructive examples, with an emphasis on pedagogical aspects.
Some important results of the last year are reviewed.
In discussing the issue of the quark-hadron duality, one of the basic
ingredients of the theory, I prove that the operator product
expansion {\em per se} is an asymptotic expansion. The behavior of
the
high order terms in this expansion determines the onset of duality
and the accuracy of the duality relations. The factorial divergence
of the high-order terms in OPE implies a sophisticated analytical
structure in the $\alpha_s$ plane, with terms of the type
$\exp [-\exp (1/\alpha_s )]$.

\end{titlepage}
\addtocounter{footnote}{-1}

\section{Introduction}

Although the weak inclusive decays of heavy flavor hadrons are
driven by weak
interactions the theory of pre-asymptotic effects is an important
applied
branch of QCD. The theory has been gradually crystallizing from the
eighties,
with a decisive breakthrough achieved  recently, in the last two or
three
years. I will try to summarize what we know and what has to be
done in future.

My talk is preceded by that of Dr. Neubert covering, in part, the
same topic.
Therefore,
I will avoid those aspects that have been already discussed. My task
is two-fold. First, I will sketch a general picture
of the phenomena lying behind the pre-asymptotic corrections.
It is remarkable that the most essential aspects of the theory
that I am going to describe
have very close parallels with the well-known phenomena from
general physics and/or quantum mechanics, for instance, the Doppler
effect, the M\"{o}ssbauer effect, etc. It is a pleasure to reveal these
analogies by translating from the language of the QCD practitioners
to that of general physics.

As it often happens, practical needs make us think of deep
theoretical questions. One of such issues, closely related to the theory
of preasymptotic effects, is the onset of duality in QCD.  When and
how it sets in? The answer to these questions is important for
understanding the accuracy of numerous predictions of our theory.
Some ideas relevant to this issue will be discussed in the end of
my talk.

The second task is to present the theory, as
it exists now, in a historical perspective. The majority of the theorists
active in the field are so young that they perceive 10-year old
results
as an ancient history. Therefore it would be in order to remind
that both, the ideas we rely upon now and the relevant mathematical
apparatus, have emerged in the eighties being, in turn,
a natural continuation of the previous work.

\section{Formulation of the problem}

Heavy flavor hadrons $H_Q$ contain a heavy quark $Q$ (practically
one
can speak of the $b$ quark and, to a lesser extent,
of the charmed quark) plus a light cloud built from
light quarks (antiquarks) and gluons. The heavy quark
$Q$ experiences a weak transition. The nature
of this transition is of no concern to us here.
It can be a radiative transition, like $b\ra
s\gamma$, semileptonic decay like $b\ra
cl\nu_l$ or a non-leptonic decay to lighter quarks, e.g.
$b\ra c\bar u d$ (Fig. 1).
It is assumed that at short distances the amplitude is
known from the electroweak theory. The task is to calculate
the decay rate of the hadron $H_Q$ and other decay characteristics:
the energy spectra, the average invariant mass of the hadronic state
produced, etc.

Asymptotically, if the heavy quark mass $m_Q$ is infinitely large,
the presence of the light cloud around $Q$ plays no role. One can
treat $Q$
as a free heavy quark decaying into free
final quarks. This is the famous parton approximation
 of the pre-historic era, the seventies. Having accepted this
prescription we, of course, immediately calculate whatever we
are asked to. Moreover, the result is obviously universal,
 independent of the structure of the light cloud. In particular,
 all lifetimes are predicted to be the same,
\beq
\Gamma(H_Q) =\Gamma (Q) .
\eeq

While this simple approach is certainly valid in the asymptotic limit
$m_Q=\infty$,
and the universality is the feature of this limit, it is totally
unsatisfactory for down-to-earth physics of the actual heavy quarks.
Indeed, a single glance at the lifetime hierarchy in the
charmed family shows how far we are from the asymptotic
 regime. The $D^+$ meson, the most long-living
 member of the family, has the lifetime exceeding that of
$\Xi_c^0$ by more than an order of magnitude,
\beq
\tau (D^+)/\tau (\Xi_c^0)\sim 12 .
\eeq
Thus, one can hardly speak of pre-asymptotic corrections
in this case -- the ``corrections" actually totally determine
the decay pattern. In the $b$ quark family they are expected to be
more modest, but still theoretical control over the inclusive decays
is impossible without calculating pre-asymptotic effects in
a systematic manner.
Their  dynamical origin  is
quite obvious: interaction of the heavy quark $Q$ and its decay
products with
the light cloud produces a series of
terms suppressed by inverse
powers of $m_Q$. (Perturbative QCD corrections emerging from the
hard gluon exchanges and suppressed
logarithmically, by powers of $\alpha_s (m_Q)$,
can be trivially incorporated in the parton model
and will not be touched upon here.)

\section{Quarks and gluons versus general physics}

Different dynamical mechanisms generate the $1/m_Q$ corrections.
The essence of the phenomenon can be best explained by a
transparent
 example familiar to everybody from the standard courses
 on quantum mechanics and nuclear physics.
Assume we are interested in the nucleon $\beta$ decay; the nucleon
 at hand is not free but is rather bound inside a nucleus.
The nucleus contains several protons and neutrons glued by
the nuclear forces and is surrounded by an electron shell (Fig. 2).
The energy release in the $\beta$ decay, $\Delta E$, is certainly
much larger
than a typical binding energy of the electrons in the shell.
Moreover, we will further assume -- counterfactually --
that $\Delta E \gg \epsilon_{nucl}$ where $\epsilon_{nucl}$
is the nucleon binding energy in the nucleus. Such nuclei do not exist
in nature, but for our pedagogical purposes it is convenient
to pretend that they do.

We start from a free neutron $\beta$ decay,
$n\ra p e\nu_e$,  and then ask what
distinguishes it from that of a bound nucleon. The answer is
well-known:
(i) the Pauli interference; (ii) the $K$ capture (by proton);
and, finally, (iii) the Fermi motion, i.e. the momentum spread
of the decaying neutron. Let us discuss these phenomena
in turn.

The Pauli exclusion principle does not allow the electron produced
to have the same momentum as that of the electrons in the shell.
Typically
the electron momentum is of order $\Delta E$, much larger than
that of the bound electrons, and interference does not occur.
In a ``corner" of the phase space, however, the electron produced
is soft implying a strong interference. This happens rarely; the effect
is
clearly
suppressed by inverse powers of $\Delta E$.

The inverse $\beta$ decay due to the $K$ capture
is simply impossible for a free nucleon. The power suppression
here is due to the fact that the amplitude is proportional to
$\psi (0)$, the bound electron wave function at the origin.

Finally, the momentum spread of the bound nucleon --
a typical bound state effect -- results in a smearing
of the momentum distribution of the produced electron.
If the decaying neutron has the momentum parallel to that
of the electron  the latter is accelerated while
for the antiparallel configuration it is decelerated. As a
result, the total decay rate is also distorted, although the
distortion is parametrically small provided that
$\epsilon_{nucl}/\Delta E \ll 1$.

The description above is qualitative. Quantitative analysis
in the case at hand presents no difficulties. The standard
quantum-mechanical methods and the knowledge of the wave
functions (of the electrons in the shell as well as the nucleons
in the nucleus) solves the problem. For instance,
to take into account the Pauli interference we just antisymmetrize
the
total wave function with respect to the coordinates of the
electron produced and that in the shell. In this way we get the
interference term automatically. To include the Fermi motion
 the neutron momentum distribution is convoluted with the
amplitude of the free decay.

The pre-asymptotic effects in the inclusive decays of heavy flavors
are
perfectly similar in nature. The decay products
of the heavy quark can interfere
with the quarks from the light cloud. The heavy quark can capture a
light
one from the cloud. Finally, the ``Fermi motion" of $Q$ is also
there: even if $H_Q$ is at rest $Q$ has a non-vanishing
spatial momentum of order of $\Lambda_{QCD}$. What is drastically
different,
however, is the fact that the wave function approach
is useless in QCD.

Indeed, the light cloud surrounding $Q$ is a messy dynamical system
composed
of an infinite number of degrees of freedom, quark and gluon,
and any attempt to reduce its dynamics to quantum mechanics of a
finite
number of particles is doomed to failure from the very beginning.

Does this mean that no adequate theory can be developed? The
question needs
no answer since the answer has been already spelled out. A
systematic approach
does exist, but one has to pay a price. If in the quantum-mechanical
case
the effects listed above are tractable for arbitrary values of
$\Delta E$ -- the requirement $\Delta E \gg \epsilon_{nucl}$
was actually unnecessary -- in the heavy flavor
decays the theory is based on the
expansion in $\Lambda_{QCD}/m_Q$; practically only the first few
terms
are calculable. Still, even this limited progress in the
quantitative description of the inclusive decays is of  great practical
value
and leads to a remarkably rich phenomenology.

\section{Formalism. A little bit of history}

As usual in practical problems from quantum chromodynamics the
only
tool existing today for dealing (analytically) with non-perturbative
effects
is the Operator Product Expansion (OPE) \cite{wilson1}. The general
idea of OPE
was formulated by Wilson even before QCD and is very simple.
Whatever amplitude is considered all contributions to
this amplitude must be systematically sorted out into two classes:
hard contributions and soft ones. One introduces an auxiliary
parameter,
$\mu$, (normalization point); if characteristic frequencies
in the given fluctuation are higher than $\mu$ this fluctuation
 is to be ascribed
to the hard part and determines the coefficient functions. In QCD
the coefficient functions are explicitly calculable provided
that $\mu$ is chosen appropriately,
$$
\alpha_s (\mu )/\pi \ll 1.
$$
(Warning: one should not think that the coefficient functions
are determined exclusively by perturbation theory; in principle,
non-perturbative contributions may also be present, for instance,
those from small-size instantons. What is important
is the fact that the weak coupling regime is applicable to the
coefficient functions. Practically, in the vast majority of
problems, the coefficient functions are given
by perturbation theory to a good approximation. This approximate
rule
is sometimes called {\em the practical version of OPE}
\cite{novikov1}.)

Soft fluctuations are referred to the matrix
elements of local operators appearing in OPE. Their contribution
is not analytically calculable in the present-day QCD. The
corresponding
matrix elements can be parametrized, however, or extracted
from other sources by using symmetry arguments, QCD sum rules
or lattice calculations, etc. All dependence on large
parameters, such as $m_Q$, resides in the coefficient functions.

The general strategy outlined above has its peculiarities in
application to the inclusive weak decays. The standard procedure
is formulated in the euclidean domain where the distinction between
the short and large distances (high and low frequencies) causes
no doubts. Kinematics of the heavy flavor decays is essentially
Minkowskian,
however; the large expansion parameter is the energy release rather
than
the Euclidean off-shellnes. The light quarks, once  produced in the
$Q$ decay,
stay forever -- thus, formally, we have to deal with the infinite time
intervals. Large distance dynamics manifests itself in full in
the exclusive decays. At the same time, in the inclusive
description, after summing over all possible channels and certain
energy
integrations, the results for the coefficient functions
depend only on the short distance dynamics. This assertion, known
as
the quark-hadron duality, can be justified by means of
an analytic continuation in the complex plane in auxiliary momenta.
The accuracy of the quark-hadron duality is exponential
in large momenta relevant to the given problem.
I will return to this point, the exponential accuracy,
later on.

Technically, in order to launch the OPE-based program one
has to consider the so-called transition operator $\hat T$ related to
the
modulus squared of the amplitude of interest \cite{shifman1}, rather
than the decay amplitude {\em per se},
\beq
\hat T (Q\ra f\ra Q) = i\int d^4x
\{ {\cal L}_W(x), {\cal L}^\dagger_ W(0)\}_T ,
\eeq
where ${\cal L}_W$ is the short-distance weak lagrangian
governing the transition $Q\ra f$ under consideration.
The finite state $f$ can be arbitrary; Fig. 3 displays, as a
particular example, the graph pertinent to the semileptonic
transition $Q\ra q l\nu_l$ where $q$ stands for the light quark.
This graph is rather symbolic; one should keep in mind that
only the skeleton lines are depicted, and all these lines
are actually submerged in a soft gluonic medium
coming from the light cloud in $H_Q$.

Next, we use the fact that the momentum operator of the
heavy quark $Q$ contains a large $c$-number part,
\beq
{\cal P}_\mu = m_Qv_\mu + \pi_\mu
\eeq
where $v_\mu$ is the
four-velocity of the heavy {\em hadron} $H_Q$, $\pi_\mu $
is the residual momentum operator,
$$
[\pi_\mu \pi_\nu ] = i G_{\mu\nu},
$$
and $ G_{\mu\nu}$ is the gluon field strength
tensor corresponding to the ``background" gluon field in the light
cloud. The presence of the large mechanical part
in the $Q$ quark momentum operator ensures
that all lines encircled by a dashed line on Fig. 3 can be
treated as hard. This block, thus, shrinks into a point;
it determines the coefficient functions
in front of local operators built from soft lines. The
the set of the soft lines includes the gluon field, light
quark fields and those describing $Q$.

Once the general idea is understood it does not take much
effort to calculate $\hat T$ in the form of the Wilson
expansion,
\beq
\hat T =\sum_n C_n {\cal O}_n,
\eeq
where the operators ${\cal O}_n$ are ordered according to their
dimensions (or, under certain circumstances, according to their
twists). The coefficients $C_n$ are proportional
to the corresponding powers of $1/m_Q$, modulo calculable
logarithms.

At the next stage we average the transition operator over the
hadronic state $H_Q$, i.e. find $\langle H_Q |\hat T|H_Q\rangle$.
It is at this stage that the internal structure of $H_Q$ enters
in the analysis. The matrix elements  $\langle H_Q |{\cal
O}_n|H_Q\rangle$
parametrize the dynamical properties of the light cloud and
its interaction with the heavy quark $Q$. They substitute the
wave function of the quantum-mechanical treatment, the
closest substitute one can possibly have in QCD.

Finally, at the very last stage, to make contact with
measurable quantities  we take the imaginary part
of the matrix element $\langle H_Q |\hat T|H_Q\rangle$.
In the example suggested above (Fig. 3) this imaginary part
is proportional to the total semileptonic decay rate,
\beq
(M_{H_Q})^{-1}{\rm Im} \langle H_Q |\hat T|H_Q\rangle =
\Gamma_{SL} .
\eeq
If the lepton energy is not integrated over but is held
fixed the imaginary part of the transition operator is proportional to
the energy spectrum, etc.

Analysis of the weak inclusive decays along these lines has been
suggested in the mid-eighties when some operators in the
expansion (5) relevant to the total lifetimes in the families of charm
and beauty have been considered \cite{shifman1}. Many elements of
what
is now called the Heavy Quark Effective Theory (HQET)
\cite{HQET,HQETr} can be found
in these
works. Further progress has been deferred
for a few years by an observation of certain difficulties in
constructing
OPE in the Minkowski domain \cite{voloshin1}. The difficulties
turned out to be
illusory
and are explained by the fact that  contributions discussed
in Ref. \cite{voloshin1}   were not fully inclusive. Once this has been
realized \cite{bigi1}
the last obstacle disappeared and a
breakthrough in the theory of pre-asymptotic effects became
inevitable.
First systematic results -- the leading
$1/m_Q^2$ corrections to the parton-model predictions --
have been obtained shortly after \cite{bigi2,blok1}. While the works
of the eighties
passed essentially unnoticed fusion of similar ideas,
which took place, independently,  in the
very beginning of the nineties, resulted in HQET in its present
formulation \cite{HQET,HQETr}.
HQET provided a simple and convenient language for discussing
the $1/m_Q$ expansions, in particular, in connection with the
semileptonic decays. General aspects of the HQET/OPE approach to
the semileptonic inclusive decays of heavy flavors have been
discussed in Ref. \cite{chay1}.   The authors of this important work
were not
familiar with
the previous development.

The current state of the theory of pre-asymptotic effects is
characterized by a rapid expansion in the practical spheres. The
$1/m_Q$, $1/m_Q^2$ and even $1/m_Q^3$ corrections are found for
probably all conceivable quantities measurable in the inclusive
decays of $H_b$ and $H_c$. An  incomplete list of
works
published in the last year and  devoted to this subject
includes a dozen of  papers [11-22]. What remains to be done to
complete
the picture? The initial excitement, quite natural
in time of the explosion of the theoretical predictions based on first
principles, should give place to an every-day routine of
extracting reliable and model-independent
numbers from experimental data. This aspect -- relation to
experiment  -- will not be touched upon in my talk.
Instead, I will demonstrate in more detail how the theory works
in two problems which seem to be most instructive.

\section{Total semileptonic widths. CGG/BUV theorem}

The inclusive semileptonic decay rate is probably the simplest and
the cleanest example. Although the parton-model analysis
(including the first radiative QCD correction) is known for more than
a decade \cite{cabibbo1} the leading non-perturbative correction was
calculated
only in 1992, a pilot project \cite{bigi2}  which opened the modern
stage and was
supplemented later by calculations of the energy spectrum
and other distributions.

If one neglects the
hard gluon exchanges, the transition operator is given by the
diagram depicted on Fig. 4. The weak Lagrangian responsible for the
semileptonic decay has the form
\beq
{\cal L} =\frac{G_F}{\sqrt{2}}V_{Qq}(\bar q \Gamma_\mu Q)
(\bar l\Gamma_\mu\nu ) ,
\eeq
where $l$ is a charged lepton, electron for definiteness,
and we assume for simplicity that the final quark $q$ is massless
($\Gamma_\mu =\gamma_\mu (1+\gamma_5)$). It is understood
that the quark lines on Fig. 4 describe propagation in a soft
gluon medium. The leptons are, of course, decoupled from gluons,
and their lines are always treated as free.

The operator product expansion for the transition operator $\hat T$
starts from the lowest-dimension operator, $\bar Q Q$ in the case at
hand.
In calculating the coefficient of this operator one
disregards any couplings of the quark lines to the {\em soft}
gluons from the cloud. The full perturbation theory resides in this
coefficient, but -- as I will explain shortly -- it is wrong
to say that the  $\bar Q Q$  term in OPE just reproduces the
parton-model
prediction for $\Gamma_{SL}$; actually it gives more.

There are no gauge and Lorentz invariant operators of dimension
four \cite{chay1} (and only Lorentz and gauge invariant operators
can
appear in the expansion for $\Gamma_{SL}$), and the first
subleading
term in OPE is due to the dimension 5 operator
\beq
{\cal O}_G =\bar Q \frac{i}{2}\sigma_{\mu\nu}G_{\mu\nu} Q.
\eeq
This is the only Lorentz scalar dimension-5 operator bilinear
in $Q$.

The corresponding coefficient $C_G$
 can be calculated in a variety of ways,
of course; the easiest one is based on the background field technique
and
the Fock-Schwinger gauge \cite{shifman3}. In this gauge the line
corresponding
to the final quark $q$ remains free (provided that we limit ourselves
to the operator ${\cal O}_G$), and the only reason why the
dimension-5
operator appears at all is the fact that we use the exact Dirac
equations of motion with respect to the heavy quark lines
$Q$ and $\bar Q$.

For those who are not familiar with the Fock-Schwinger gauge
technique
let me add a remark elucidating the above statement. Explicit
calculation of
the diagram of Fig. 4, where all quark lines are considered in the
background gluon field (with the Fock-Schwinger gauge
condition), yields the operator $\bar Q(0) \not\!\!{p} (p^2)^2Q(0)$
where $p =i\partial$. Now, we would like to rewrite
$p$ in terms of the covariant derivative ${\cal P} = p +A$ and
the gluon field strength tensor -- which is perfectly trivial in
the Fock-Schwinger gauge -- and then take advantage of the fact that
 $\not\!\!{\cal P} Q = m_Q Q$. It is not difficult to see that
in the linear in $G_{\mu\nu}$ approximation one can substitute
$$
\bar Q(0) \not\!\!{p} (p^2)^2Q(0)
\ra \bar Q(0) \not\!\!{\cal P} ({\cal P}^2)^2Q(0) .
$$
Supplementing this expression by the equalities
$$
{\cal P}^2 = {\not\!\!{\cal P}}^2 -\frac{i}{2}\sigma G\,\, {\rm and}\,\,
\not\!\!{\cal P} Q = m_Q Q
$$
we conclude that in the Fock-Schwinger gauge the operator
$\bar Q\not\!\!{p} (p^2)^2Q$ reduces to $m_Q^5\bar QQ -2m_Q^3 {\cal
O}_G$.
In other words, calculating the diagram of Fig. 4 with
free fermions we get, for free, both coefficients --
the coefficient in front of $\bar Q Q$ and that in front of
${\cal O}_G$!

I bring my apologies for this brief technical digression intended
to demonstrate the beauty of the Fock-Schwinger technique
which, unfortunately, is still unfamiliar to many heavy quark
practitioners. Now it is time to present the final prediction
\cite{bigi2}
for the inclusive semileptonic decay rate,
$$
\Gamma (H_Q\ra X_q l\nu_l) =
\frac{G_F^2m_Q^5|V_{Qq}|^2}{192\pi^3}
\times
$$
\beq
\left\{ \frac{\langle H_Q|\bar QQ|H_Q\rangle}{2M_{H_Q}}
-\frac{2}{m_Q^2}\frac{\langle H_Q|{\cal O}_G|H_Q\rangle}{2M_{H_Q}}
+ {\cal O}(m_Q^{-3})\right\} .
\eeq
Perturbative hard gluon corrections are omitted here.

This is almost the end of the story. To complete the analysis one
needs to know
the matrix elements of the operators $\bar Q Q$ and ${\cal O}_G$
over the
hadronic state $H_Q$ of interest. These matrix elements, clearly,
depend
on the structure of the light cloud which is beyond our computational
abilities
at present. Fortunately, they are known independently!

Indeed, ${\cal O}_G$ is the lowest-dimension operator responsible
for the  spin splittings -- the mass difference between the vector
and pseudoscalar mesons (e.g. $B^*$ and $B$). Thus,
\beq
\frac{\langle B|{\cal O}_G|B\rangle}{2M_B} \equiv \mu_G^2
=\frac{3}{4}(M_*^2-M^2)
\eeq
where $M_*$ and $M$ are the $B^*$ and $B$ masses, respectively.
The
right-hand side of eq. (10)
is well measured experimentally. As for $\langle H_Q|\bar
QQ|H_Q\rangle$,
let us compare the operator $\bar QQ$ with the current
$\bar Q\gamma_0 Q$ whose matrix element over any
$Q$ flavored hadron is unity, of course (in the $H_Q$ rest frame). We
are lucky
again. The difference $\bar QQ - \bar Q\gamma_0 Q$ is due to the
lower
components of the bispinors, and for heavy quarks these are
reducible to the upper components by means of the
equations of motion expanded in $1/m_Q$. In this way we arrive at
\cite{bigi2}
\beq
\bar QQ = \bar Q\gamma_0Q +\frac{1}{2m_Q^2}{\cal O}_G
-\frac{1}{2m_Q^2}{\cal O}_\pi +{\cal O}(m_Q^{-4}) ,
\eeq
where
$$
{\cal O}_\pi =\bar Q {\vec\pi}^2 Q
$$
has the meaning of the average spatial momentum squared
of the heavy quark $Q$ inside $H_Q$.

As you probably remember, I told you that all perturbation
theory is hidden in the coefficient of the operator
$\bar QQ$. The parton-model result is recovered if one
approximates the matrix element $(2M_{H_Q})^{-1}
\langle H_Q|\bar QQ|H_Q\rangle$ by unity. Now we
see that this is actually the zeroth order approximation;
the very same operator $\bar QQ$ gives rise
to $1/m_Q^2$ and higher-order terms, for instance,
\beq
\frac{1}{2M_B}\langle B|\bar bb|B\rangle
= 1 +\frac{\mu_G^2}{2m_b^2} -
 \frac{\mu_\pi^2}{2m_b^2} +...
\eeq
where $\mu_\pi^2$ parametrizes the matrix element of ${\cal
O}_\pi$,
$$
\mu_\pi^2 = (2M_{H_Q})^{-1}\langle H_Q|\bar
Q{\vec\pi}^2Q|H_Q\rangle
{}.
$$

Assembling all these results together we finally find \cite{bigi2}
$$
\Gamma (B\ra X_u l\nu_l) =
\frac{G_F^2m_Q^5|V_{bu}|^2}{192\pi^3}
\times
$$
\beq
\left\{ 1+ \,{\rm zero}\times\frac{1}{m_b}\, +
\left[ \frac{1}{2m_b^2}(\mu_G^2 - \mu_\pi^2)\right] -
\left[ \frac{2}{m_b^2}\mu_G^2\right]+...\right\} .
\eeq
where the expression in the first square brackets is due to
the $1/m_b^2$ correction in the matrix element of $\bar bb$,
the term in the second square bracket comes from
the operator ${\cal O}_G$ in OPE for $\Gamma_{SL}$, and the
dots denote higher-order terms in $1/m_b$.

The $\mu_\pi^2$ part in the first square bracket has a very
transparent
physical interpretation. This is nothing else than the quadratic
Doppler effect reflecting the time dilation for the heavy
quark decaying in flight. Indeed, in the rest frame
of the decaying hadron the quark $b$ has a spatial momentum
prolonging its lifetime (diminishing $\Gamma$).
The effect linear in this momentum
drops out because of averaging over angles, while the
quadratic effect survives. It is quite obvious, from
the first glance, that the coefficient of $\mu_\pi^2$ in eq.
(13) is exactly as it is expected to be in this picture.

A remarkable fact which I have especially emphasized in
eq. (13) is the absence of the correction linear in
$1/m_Q$ in the total semileptonic width.
This practically important {\em theorem} (CGG/BUV theorem)
is derived from two components -- the absence of the dimension-4
operators
in OPE and the absence of $1/m_Q$ corrections
in the matrix elements of the operator $\bar QQ$.
The first circumstance has been noted in Ref. 10. The authors of this
work believed, however, that linear in $1/m_Q$ terms
may appear in $(2M)^{-1}\langle H_Q|\bar QQ|H_Q\rangle $.
The non-renormalization theorem for
$(2M)^{-1}\langle H_Q|\bar QQ|H_Q\rangle $ at the $1/m_Q$ level
has been established in Ref. 8 completing the proof of the theorem.

\section{The impact of the heavy quark motion in $H_Q$}

Even if the $Q$ flavored hadron is nailed at the origin so that
its velocity vanishes the heavy quark $Q$ moves inside
the light cloud, its momentum being of order $\Lam$. This is
the QCD analog of the Fermi motion of the nucleons in the nuclei.
It is quite clear that this motion affects the decay spectra.
Say, if the ``primordial" heavy quark momentum is parallel to that of
the lepton pair, the pair gets boosted. A distortion of spectra may
lead
to corrections in the decay rate. It is equally clear that
this effect is pre-asymptotic (suppressed
by inverse powers of $m_Q$): while typical energies of the
decay products are of order $m_Q$ a shift due to the heavy quark
motion is of order $\Lam$.

A few models engineered to describe this effect at a purely
phenomenological
level exist in the literature for years (e.g.  Ref. 25). Now one can
finally
approach the issue scientifically, backed up by
a solid QCD-based analysis.
The theory of this phenomenon is very elegant. I would like to
demonstrate
basic points in a simplest example, $Q\ra q\gamma$ transition.
Moreover, to avoid inessential technicalities (which can be
readily worked out, though) I will neglect the quark and photon
spins,
assuming that both fields, $Q$ and $q$ are spinless. As for the
mass of the final quark $m_q$ it  will be treated as a free
parameter which can vary from zero almost up to $m_Q$.
(For our approach to be valid we still need
that $\Delta m \equiv m_Q-m_q\gg\Lam$ although the
mass difference may be small compared to the quark masses.
As a matter of fact, this {\em small velocity} (or SV) limit
\cite{voloshin2}
is very interesting and instructive, and I will return
to its discussion later on.)

Thus, for pedagogical purposes we will consider, following Ref. 17, a
toy
model,
\beq
{\cal L}_\phi =h\bar Q\phi q +\,\, {\rm h.c.}
\eeq
where $\phi$ is a toy ``photon" field and $h$ is a coupling
constant. To begin with,
it is convenient to put the final quark mass to zero. At the level of
the
free quark decay the photon energy is fixed by the two-body
kinematics, $E_\phi =m_Q/2$. In other words, the photon energy
spectrum
is a monochromatic line at $E_\phi =m_Q/2$ (Fig. 5). On the other
hand,
in the actual hadronic decays
the kinematical boundary of the spectrum lies at
$M_{H_Q}/2$. Moreover, due to multiparticle final states
(which are, of course, present at the level of the hadronic decays) the
line will be smeared. In particular, the {\em window} --
a gap between $m_Q/2$ and $M_{H_Q}/2$ -- will be closed
(Fig. 6). There are two mechanisms smearing the monochromatic
line of Fig. 5. The first is purely perturbative: the final quark $q$
can shake off a hard gluon, thus leading to  the three-body
kinematics. The second mechanism is due to the ``primordial"
motion of the heavy quark $Q$ inside $H_Q$ and is
non-perturbative. Only the second
mechanism will be of interest for us here. The theory of the
line shape in QCD resembles that of the M\"{o}ssbauer
effect.

Thus, the  question to be answered is: ``how can one translate
an intuitive picture of the $Q$ quark primordial motion inside $H_Q$
in QCD-based theoretical predictions, not just plausible models?"
The OPE/HQET approach is applicable here in full, with
a few additional -- rather transparent -- technical elements.

The basic object of consideration is again the transition operator (3)
(with ${\cal L}_W$ substituted by ${\cal L}_\phi$). Since we are
interested in
the energy spectrum the ``photon" momentum $q$ must be fixed.

Then in the approximation of Fig. 7 (no hard gluon exchanges)
the transition operator takes the form
\beq
\hat T = -\int d^4 x (x|\bar Q\frac{1}{(P_0 -q +\pi)^2}Q|0) ,
\eeq
where the background field technique is implied and
$(P_0)_\mu=m_Qv_\mu$. I remind the toy model with no spins for
the quark and photon fields is taken for illustration; bellow a
convenient
shorthand notation
$$
k=P_0-q = m_Qv-q
$$
will be used.

If $\hat T$ is calculated in the operator product expansion
the problem is solved since
\beq
\frac{d\Gamma}{dE} =
\frac{h^2E}{4\pi^2M_{H_Q}}\Im
\langle H_Q|\hat T |H_Q\rangle .
\eeq

To construct OPE we observe that, as in the previous problem
of the total semileptonic widths, the momentum operator
$\pi$ corresponding to the residual motion of the heavy quark
is $\sim\Lam$ and the expansion in $\pi /k$ is possible.
Unlike the problem of the total widths, however, now in the
interesting
energy range $k^2$ is anomalously small, the expansion parameter is
of order unity, and an infinite series of terms has to be summed up.

To elucidate this statement let us examine different terms
in the denominator of the propagator,
$$
k^2 +2\pi k +\pi^2 .
$$
The effect of the primordial motion plays a key role in the window
and in the adjacent region below it,
\beq
|E-(1/2)m_Q|\sim \La
\eeq
where
$
\La = M_{H_Q}-m_Q .
$
It is quite trivial to find that in this domain
$$
k_0\sim|\vec k|\sim m_Q/2, \,\, k^2\sim m_Q\La ;
$$
in particular, at the kinematical boundary (for the maximal value
of the ``photon" energy)
$k_0 = M_Q/2$ and $k^2 = -m_Q\La$. Hence, inside the line
$$
k^2\sim k\pi\gg \pi^2.
$$
In other words, when one expands the propagator of the final quark
in eq. (15)
in $\pi$, in the leading approximation all terms $(2k\pi /k^2)^n$
must be
taken into account while terms containing $\pi^2$ can be omitted.
The first subleading correction would contain one $\pi^2$ and
arbitrary number
of $2k\pi$'s, etc. For those who still remember the
theory of deep inelastic scattering (DIS) of the late sixties and early
seventies this may sound suspiciously familiar.
Yes, you are right; the theory of the line shape in the radiative
transition of the heavy  quark  into a massless one
can be viewed as a version of the standard theory of DIS.

Thus, in this problem it is twist of the operators in OPE,
not their dimension, that counts. Keeping only those terms
in the expansion that do not vanish in the limit $m_Q\ra\infty$
(analogs of the twist-2 operators in DIS)
we get the following series for the transition
operator \cite{bigi4}
\beq
\hat T = -\frac{1}{k^2}\sum_{n=0}^{\infty}\left(
-\frac{2}{k^2}\right)^n
k^{\mu_1}...k^{\mu_n} \left( \bar Q\pi_{\mu_1}...\pi_{\mu_n}Q
- {\rm traces}\right) .
\eeq
Traces are subtracted by hand since they are irrelevant
anyway; their contribution is suppressed as
$k^2\pi^2 /(k\pi )^2\sim\La /m_Q$ to a
positive power. Another way to make
the same statement is to say that in eq. (18) the four-vector
$k$ can be considered as {\em light-like}, $k^2=0$.

After the transition operator is built the next step is averaging
of $\hat T$ over the hadronic state $H_Q$. Using only the
general arguments of the Lorentz covariance one can write
\beq
\langle H_Q|\bar Q\pi_{\mu_1}...\pi_{\mu_n}Q - {\rm
traces}|H_Q\rangle
=a_n\La^n (v_{\mu_1}...v_{\mu_n}  - {\rm traces})
\eeq
where $a_n$ are constants parametrizing the matrix elements.
Their physical meaning will become clear momentarily. Right now it
is worth
noting that the term with $n=1$ drops out ($a_1=0$). Indeed,
$\langle H_Q| \bar Q\vec\pi Q|H_Q\rangle$ is obviously zero for
spinless $H_Q$ while $\pi_0$ through the equation of motion
reduces to ${\vec\pi}^2/(2m_Q)$ and is of the next
order in $1/m_Q$. Disappearance of $\bar Q\pi_\mu Q$ means
that there is a gap in dimensions of the relevant operators.

Let us write $a_n$'s as moments of some function $F(x)$,
\beq
a_n=\int dx x^n F(x) .
\eeq
Then, $F(x)$ is nothing else than the primordial line-shape function!
(That is to say,
$F(x)$ determines the shape of the line before it is deformed
by hard gluon radiation; this latter deformation is
controllable in perturbative QCD). The variable $x$ is
related to the photon energy,
$$
x= (2/\La ) [E - (1/2)m_Q] .
$$
If this interpretation is accepted -- and I will prove
that it is correct --
it immediately implies that (i) $F(x) > 0$, (ii) the upper limit
of integration in eq. (20) is 1, (iii) $F(x)$ exponentially
falls off at negative values of $x$ so that practically
the integration domain in eq. (20) is limited from below at
$-x_0$ where $x_0$ is a positive number of order unity.

To see that the above statement is indeed valid we substitute
eqs. (19), (20)  in $\hat T$,
\beq
\langle H_Q| \hat T|H_Q\rangle
=-\frac{1}{k^2}\sum_n \int dy F(y) \, y^n
\left( -\frac{2\La kv}{k^2 +i0}\right)^n ,
\eeq
and sum up the series. The $i0$ regularization will prompt us
how to take the imaginary part at the very end. In this way we
arrive at \cite{bigi4}
\beq
\frac{d\Gamma}{dE} =-\frac{4}{\pi}
\Gamma_0 \frac{m_QE}{M_{H_Q}}\Im
\int dy F(y)
\frac{1}{k^2 + 2y\La kv +i0}
=(2/\La )\Gamma_0 F(x) ,
\eeq
where the variable $x$ is defined above and $\Gamma_0$
is the total decay width in the parton approximation.
Corrections to eq. (22) are of order $\La /m_Q$.

Thus, we succeeded in getting the desired smearing:
the monochromatic line of the parton approximation is replaced
by a finite size line whose width is of order
$\La$. The pre-asymptotic
effect we deal with is linear in $\Lam /m_Q$.

At this point you might ask me how this could possibly happen.
I told you a few minutes ago that there is a gap
in dimensions of operators in OPE -- no operators
of dimension 4 exist -- and the correction
to $\bar Q Q$ is also quadratic in $1/m_Q$ (the CGG/BUV theorem).
No miracles -- the occurrence of the effect linear
in $\Lam /m_Q$ became possible due to the summation of the
infinite
series in eq. (18); no individual term in this series gives
rise to $\Lam /m_Q$.

To avoid misunderstanding it is worth explicitly stating
that the primordial line function is {\em not} calculated;
rather $F(x)$ is related to the light-cone distribution function of the
heavy quark inside $H_Q$, {\em vis.} $\langle \bar Q(n\pi
)^nQ\rangle$,
$n^2=0$, or more explicitely
\beq
F(x)\propto\int dt{\rm e}^{ixt\La}
\langle H_Q|{\bar Q}(x=0 ){\rm e}^{-i\int_0^t nA(n\tau )d\tau}
Q(x_\mu=n_\mu\tau)
|H_Q\rangle ,
\eeq
where $n$ is a light-like vector
$$
n_\mu = (1,0,0,1).
$$
Moreover, this primordial function is not the one that will be
eventually
measured from $d\Gamma /dE$; the actual measured line shape will
be essentially deformed by radiation of the hard gluons. I will say
a few words about this effect later on.

Although we can not calculate $F(x)$  still our analysis
is useful in many respects. First of all, a few first moments
of $F(x)$ are known, as well is its qualitative behavior.
Second, the QCD-based approach tells us that
the ``scientific" distribution function is one-dimensional,
while in a naive model one would rather introduce a
three-dimensional distribution in the $Q$ spatial momentum.
The reason is obvious: the momentum operator $\pi_\mu$ does not
commute with
itself, $[\pi_\mu\pi_\nu ] = i G_{\mu\nu}$, while $n\pi$ commutes.
Finally, the logarithmic evolution of $F(x)$ is calculable in
perturbative QCD.

For real quarks and photons, with spins switched on,  calculation
of the line shape is carried out in Refs.  17,18.

\section{Varying the mass of the final quark}

So far I was discussing the transition into a massless final quark.
If we looked at the line shape presented on Fig. 5 more
attentively, through a microscope, we would notice that a
smooth curve is obtained as a result of adding up many channels,
specific decay modes.
A typical interval in $E$ that contains already enough channels
to yield a smooth curve after summation is $\sim\La^2/m_Q$.
In other words, in terms of
the photon energy the duality interval is $\sim\La^2/m_Q$.
Roughly one can say that
the spectrum of Fig. 5 covers altogether $m_Q/\La$  resonance
states poduced in the $H_Q$ decays and composed
of $q$ plus the spectator (I keep in mind here
that the final hadronic state is produced through decays
of highly excited resonances, as in the multicolor QCD). These states
span the window between $m_Q/2$ and $M_{H_Q}$ and the adjacent
domain
to the left of the maximum at  $E=m_Q/2$.

It is very interesting to trace what happens with the
line shape, the duality interval and the primordial distribution
function as the final quark mass $m_q$ increases \cite{bigi4}.

The most obvious kinematical effect is the fact that
the window shrinks. When we eventually come to the SV limit
\cite{voloshin2},
$$
\Lam\ll \Delta m \equiv m_Q-m_q\ll m_{Q,q}
$$
it shrinks to zero. Indeed, in this limit
the photon energy in the two-body quark decay,
$\Delta m (1+ \Delta m(2m_Q)^{-1}) $ differs from the
maximal photon energy in the hadronic decay,
$\Delta M (1+ \Delta M(2M_Q)^{-1}) $,  only by a tiny amount
inversely proportional to $m_Q$ ($\Delta m$ and
$\Delta M$ stand for the quark and meson mass differences,
respectively).
 Simultaneously with the
shrinkage of the window the peak becomes more asymmetric
and develops a two-component structure (Fig. 8). The dominant
component of the peak, on its right-hand side, becomes
narrower and eventually collapses into a delta
function in the SV limit. A shoulder develops on the left-hand
side; the duality interval becomes larger and the number
of the hadronic states populating the line becomes smaller.
When we approach the SV limit the delta function component
will consist of exactly one sate -- the lowest-lying
$q$ containing meson (the elastic component) while
the shoulder (the inelastic component) will include
several excitations and will stretch down to $E=
E_{max}-$ several units
$\times \La$. The light-cone distribution function will
evolve and will continuously pass into the {\em temporal}
distribution function determining the form of the
spectral line in the shoulder (Fig. 9).

This rather sophisticated picture, hardly reproducible in
the naive models, stems from the same analysis \cite{bigi4}. The
transition
operator $\hat T$ now has the form
\beq
\hat T = \frac{1}{m_q^2-k^2}\sum_{n=0}^\infty
\bar Q \left( \frac{2m_Q\pi_0 +\pi^2 - 2q\pi}{m_q^2-k^2}\right)^n Q,
\eeq
where $q$ is the $\phi$ momentum. Notice that
$2m_Q\pi_0 +\pi^2$ acting on Q yields zero (the equation of motion)
and in the SV limit $q$ must be treated as a small
parameter,
$$
q_0 /m_Q\equiv E/m_Q = v\ll 1;
$$
$v$ is the spatial velocity of
the heavy quark produced. Although $v$ is small the inclusive
description we develop is still valid provided that
$\Delta m \gg\Lam$.

In the zeroth order in $q$ the only term surviving in the
sum (24) is that with $n=0$, and we are left with the single pole,
the elastic contribution
depicted on Fig. 9. This is the extreme realization of the
quark-hadron
duality. The inclusive width is fully saturated
by a single elastic peak. What might seem to be a miracle at first
sight
has a symmetry explanation -- the phenomenon is explained by the
heavy
quark symmetry which has been noted in this particular context
in Refs.  26, 27. (The heavy quark symmetry is also called
the Isgur-Wise symmetry \cite{isgurwise}. These authors generalized
the observation of Refs. 26, 27 to arbitrary values of $v$.) On the
other hand, the fact that the parton-model
monochromatic line is a survivor of hadronization is akin
to the M\"{o}ssbauer effect.

If terms ${\cal O}(v^2)$ are switched on the transition operator
acquires an
additional part,
\beq
{\hat T}_{v^2} = \frac{4}{3}{\vec q}^2
\frac{1}{(m_q^2-k^2)^3}
\sum_{n=0}^\infty \left( \frac{2m_Q}{m_q^2-k^2}\right)^n\, \bar Q
\pi_i\pi_0^n\pi_i  Q.
\eeq
{}From this expression it is obvious that the shape of the $v^2$
shoulder
is given \cite{bigi4} by the temporal distribution function $G(x)$
whose moments
are introduced through the matrix elements
\beq
\langle H_Q |\bar Q\pi_i\pi_0^n\pi_i Q|H_Q\rangle = \La^{n+2}\int dx
x^n G(x) .
\eeq
Alternatively, $G(x)$ can be written as a Wilson line along the time
direction,
\beq
G(x)\propto \int dt{\rm e}^{ixt\La}
\langle H_Q|\bar Q (t=0,\vec x =0)
\pi_i {\rm e}^{-i\int_0^t A_0(\tau )d\tau}\pi_i Q(t, \vec x
=0)|H_Q\rangle .
\eeq

Intuitively it is quite clear why the light-cone
distribution function gives place to the temporal
one in the SV limit. Indeed, if the massless final quark propagates
along
the light-cone, for $\Delta m \ll m$  the quark $q$ is at rest in the
rest frame
of $Q$, i.e. propagates only in time.

In terms of $G(x)$ our prediction
for the line shape following from eq. (25) takes the form
$$
\frac{d\Gamma}{dE}\propto
\left[ 1-\frac{v^2}{3}\int\left(\frac{1}{y^2}+\frac{\La
/E_{max}}{y}\right)
G(y)dy\right]\delta (x)+
$$
\beq
\frac{v^2}{3}\left(\frac{1}{x^2}+\frac{\La /E_{max}}{x}\right) G(x) ,
\eeq
where $x=(E-E_{max})/\La$. The $v^2$ corrections affect both, the
elastic peak
(they reduce the height of the peak) and the shoulder (they create
the
shoulder). The total decay rate stays intact, however: the
suppression of the elastic peak is compensated by the integral over
the inelastic contributions in the shoulder. This is the Bjorken sum
rule \cite{bjorken1,bjorken2}.
 It is important that we do not have to guess  or make {\em ad hoc}
assumptions --
 a situation typical for model-building -- the theory itself tells us
what distribution function enters in this or that case and in what
particular way.

\section{Inclusive semileptonic decays$^{[16,17,19,21]}$}

The very same distribution functions that determine the line
shape in the radiative transitions appear in the problem of the
spectra in the semileptonic decays. In particular,
in $b\ra ul\nu$ we deal with $F(x)$. Their analysis
is absolutely crucial if one addresses the behavior in the so-called
end-point region (i.e. the region inaccessible in
the parton model and close to it, an analog of the {\em window} in
the radiative
transitions).

Apart from trivial kinematical modifications --
for instance, the occurrence of several structure functions --
the only change is the expression for the variable $x$. More exactly,
as was shown above, for {\em massless} final quark
in both cases, radiative transitions
and semileptonic decays,
 $$x=-k^2/(2\La kv),$$
but in the latter case
$q^2 >0$ while in the radiative transitions $q^2=0$,
and this seemingly insignificant difference is practically important.
Indeed,
if $q^2=0$ the notion of scaling makes no sense -- kinematically
there
is only one variable, the photon energy E, and $x\propto (E-
E_{max})$.
On the other hand, in the inclusive decays the structure functions
depend,
generally speaking, on two independent variables,
$k^2$ and $kv$ (or $q^2$ and $q_0$). The fact  that the structure
functions in $B\ra X_u l\nu$ are actually functions of  a single
combination $x$ is a very strong statement fully equivalent to the
Bjorken scaling in deep inelastic scattering. Guesses about
a scaling behavior in the inclusive semileptonic
decays are known in the literature \cite{paschos}. This is the first
time ever
we are able to say for sure what sort of scaling takes place,
where it is expected to hold and where and how it will be violated.

In the most naive parton model, when gluons are not considered at
all,
$k^2=0$. The primordial heavy quark motion smears $k^2$ so that
actually
\beq
k^2\sim\La k_0.
\eeq
I will sometimes refer to this domain as to ``primordial".
In this domain the structure functions depend on
\beq
x = -\La^{-1}k^2/2k_0 = -\La^{-1}k^2/(k_0+|\vec k|) = -\La^{-1}(k_0-
|\vec k|)
\eeq
where in the denominator
the difference between $k_0$ and $|\vec k|$ is neglected which is
perfectly
legitimate in the primordial domain (29). In the rest of the
phase space this substitution is wrong, of course, but there
the above scaling is not going to be valid anyway, so there is no need
to
bother. From eq. (30) we see that in the primordial  domain
the structure functions depend on a single light-cone combination
$$q_0+|\vec q|,$$
 rather than on $q_0$ and $q^2$ separately.

The primordial  structure functions fall off -- presumably
exponentially -- outside the primordial domain.
The hard gluon emissions will populate the phase space outside this
domain creating long logarithmic tails. The primordial part is buried
under these tails. Therefore, outside the primordial
domain one can not expect that the structure functions depend
on the single combination $q_0+|\vec q|$.

Perturbative corrections as well
as those due to higher twist operators violate scaling in the
primordial domain as well. If the latter are generically small
the perturbative corrections in the end-point region,
due to the Sudakov double log suppression, are expected
to be large, so that practically in the measured structure functions
the scaling feature will be implicit. Needless to say
that perturbative violations of scaling are calculable.

I will not go into further details which are certainly
important if one addresses the problem of extraction of $V_{ub}$
from experimental data.  Some of them are discussed
in the current literature \cite{bigi4,falk1}, others still have to be
worked out.
Applications of
the theory to data analysis is a separate topic going beyond the scope
of the present talk. The only obvious remark on the visibility
of the primordial motion in experimental data seems to be in
order: the electron energy spectrum in
$B\ra X_ue\nu$ is probably the worst place for studying
$F(x)$. Indeed, here the primordial distribution function
enters in an integrated form even in the absence of the hard
gluon corrections. A smearing of $F(x)$ is unavoidable if we
integrate over the neutrino variables, and then
the effect of $F(x)$ is diluted, and is seen to a much
lesser extent than in the radiative transitions. In the
inclusive semileptonic decays a much better place to search
is the double differential distribution in $q_0$ and $q^2$.

\section{$b\ra c l \nu$}

The formalism described above is fully applicable
\cite{bigi4,mannel2},
in principle, to the semileptonic inclusive transitions $b\ra c l \nu$
where the final quark $c$ can be also treated as heavy, although at
the same time , $m_c^2\ll m_b^2$. The ratio $m_c^2/m_b^2
\approx 0.07$ is a small parameter while $m_c^2/(\La m_b)
\sim 1$.  Technically, the situation becomes much more
sophisticated, however, since the type of the distribution function
reflecting the primordial motion of the $b$ quark inside the $B$
meson will depend now on the value of $q^2$ where $q$ is the
total momentum of the lepton pair.

If $q^2$ is small, $q^2< m_c^2$, one recovers \cite{mannel2} the same
light-cone
function $F(x)$ as in the transition $b\ra ul\nu$ or $b\ra s\gamma$.
Modifications  are  marginal. First, some extra terms
explicitly proportional to $m_c/m_b$ are generated in the structure
functions due to the fact that
$\not\!\!{\cal P} +m_c$ replaces $\not\!\!{\cal P}$
in the numerator of the quark Green function. Moreover,
if in the $b\ra ul\nu$ transitions the scaling variable in the
end-point domain is
\beq
x=\La^{-1}(q_0 +|\vec q| -m_b) ,
\eeq
in the $b\ra cl\nu$ transition it is shifted by a constant term of
order 1,
\beq
x=\La^{-1}(q_0 +|\vec q| -m_b) +\frac{m_c^2}{\La m_b} .
\eeq
Finally, subleading (higher-twist) terms in $b\ra ul\nu$
generically give rise to corrections of order $k^2/k_0^2\sim\La
/m_b$. In the beauty-to-charm decays these corrections are
${\cal O}((\La m_b +m_c^2)/m_b^2)$.

As $q^2$ increases the corrections to the description \cite{mannel2}
based on the light-cone function
grow and eventually
blow up when $q^2$ approaches its maximal value,
$$
q^2_{max} = (M_B-M_D)^2 ,
$$
since here $k_0\sim m_c$, not $m_b$. Practically this happens at
\beq
\sqrt{q^2} > M_B-2M_D .
\eeq
The domain (33) corresponds to the SV regime which I have already
discussed, with fascination, in the toy example above. In the SV
limit the velocity of $H_c$ produced is small,
$$
|\vec v|^2 \approx \frac{(M_B-M_D)^2 -q^2}{M_BM_D}\ll 1,
$$
and the light-cone distribution function becomes irrelevant.
The primordial motion of the $b$ quark is described \cite{bigi4} in
this regime
by the temporal distribution function $G(x)$ where
$$
x\approx \La^{-1}(q_0 -\Delta m),
$$
$\Delta m $ is the quark mass difference coinciding, to the
leading order with $M_B-M_D$.
The "corner" of the phase space where
we find ourselves close to the SV limit is responsible for roughly
1/2 of all events in $B\ra X_cl\nu$.

Thus, changing $q^2$ from zero to $q^2_{max}$ results in an
evolution of the distribution function appearing in theoretical
formulae for $d\Gamma (B\ra X_cl\nu)$, from
light-cone to  temporal, through a series of intermediate
distributions. The
physical reason for this evolution is quite clear -- what distribution
function is actually measured depends on the parton-model velocity
of the quark produced in the $b$ decay.

In calculating the lepton energy spectrum,
$d\Gamma /dE_l$,  one needs to integrate over a range of $q^2$.
In doing so, we inevitably smear all these evolutionary stages.
How exactly the effect of the primordial motion in
$d\Gamma (B\ra X_cl\nu)/dE_l$ looks like -- nobody knows at
present.  An attempt to express this effect in terms of the light-cone
function $F(x)$ has been recently reported \cite{mannel2}. One
should expect
that, given the actual values of the quark masses, the expression for
$d\Gamma (B\ra X_cl\nu)/dE_l$
obtained in this work will have corrections ${\cal O}(1)$.

\section{The Roman distribution function}

The moments of the light-cone distribution function
$F(x)$ are related to the matrix elements of the operators
(19). This infinite set of matrix elements
codes information about the bound-state structure of
$H_Q$. Needless to say that we are not able to calculate this infinite
set in the present-day QCD; our possibilities are limited
to a few first moments. Since the desire to get an idea
of the primordial distribution function, to visualize $F(x)$,
is naturally strong it seems reasonable to adopt a pragmatic attitude.
Let us try to conjecture a qualitatively acceptable function,
whose properties are compatible with the expected features of the
primordial distribution function. This conjectured function will
depend on several free parameters to be fixed from the
moments we know.

No extensive guesswork is needed. Surprisingly, a phenomenological
model existing for
more than a decade will provide us with a quite adequate input for
$F(x)$. I mean the so-called AC$^2$M$^2$ model \cite{altarelli1} for
the ``Fermi"
motion
of the heavy quark inside $H_Q$.

The physical essence of the model is very simple. It assumes that,
say, the $B$
meson is a compound system consisting of the heavy quark $b$ and
a spectator with a fixed mass $m_{sp}$ (one of two free parameters
of the model). The spectator component has a momentum
distribution
$\Phi (|p|)$ where $\vec p$ is a three-dimensional momentum of the
spectator. The energy of the $b$ quark is
$E_b=M_B-(|\vec p |^2+m_{sp}^2)^{1/2}$. The mass of the $b$
quark then has to be equal to $E_b$ through order $p$. Thus, we
arrive
at the notion of the floating $b$ quark mass
\beq
m_b^f \approx  M_B-(|\vec p |^2+m_{sp}^2)^{1/2},
\eeq
technically a most important ingredient of the  AC$^2$M$^2$ model.
The $b$ quark with momentum $-\vec p$ decays into
$ue\nu$ in flight; after averaging over
$$
\Phi (|\vec p|) =\frac{4}{p_F^3\sqrt{\pi}}{\rm e}^{-\frac{p^2}{p_F^2}}
$$
we obtain
all relevant decay spectra (here $p_F$, a Fermi momentum, is a
second
free parameter of the model).

This model is routinely used for years in analyzing the experimental
data on the inclusive semileptonic decays of $B$'s. Given all
its na\"{i}vet\`{e} it is quite remarkable that
it still reproduces essential features of the QCD-based theory
of pre-asymptotic corrections. First of all, it has been
demonstrated \cite{randall1,bigi5} that the corrections of the first
order
in $1/m_b$ to the total decay rates are absent in this model,
the CGG/BUV theorem. Second, in the primordial domain the correct
scaling is recovered: the structure functions do indeed depend
only on the combination (30) or (31).

It is not difficult to extract \cite{bigi5} the light-cone
distribution function stemming from the  AC$^2$M$^2$ model,
\beq
F_{Rom}(x) =\frac{1}{\sqrt{\pi}}\,\frac{\La}{p_F}\,
{\rm exp}
\left\{ - \frac{1}{4}\left[
\frac{p_F}{\La}\frac{\rho}{1-x}
-\frac{\La}{p_F}(1-x)\right]^2\right\} ,
\eeq
where $Rom$ stands for Roman, and
$$
\rho = (m_{sp}/p_F)^2 .
$$
The vanishing of the first moment implies
\beq
\frac{\La}{p_F} = \frac{\rho}{\sqrt{\pi}} {\rm e}^{\rho
/2}K_1(\frac{\rho}{2}) ,
\eeq
where $K_1$ is the McDonald function.
The  Roman function satisfies the requirements one
would expect from  the primordial distribution function
in the inclusive decays into massless quarks. First,
it vanishes when one approaches $x=1$ from below, so that no
spectrum is generated beyond the physical kinematical boundary.
Moreover, at the negative values of $x$ it falls off exponentially
once $|x| \gg 1$.

The Roman function depends on two parameters to be fitted,
$m_{sp}$ and $p_F$. Usually they are fitted from experimental data.
Previously, when the theory of pre-asymptotic effects did not
exist, there was no clear idea how to fit the data properly.
New understanding calls for a new analysis, currently
under way. Meanwhile it is instructive to see
what can be said theoretically.

It is not difficult to show that the AC$^2$M$^2$ model constrains the
second moment from above, $\langle x^2\rangle <(\pi -2)/2$. The
largest
possible value, $\langle x^2\rangle \sim 0.6$ is achieved if $\rho
\ra 0$. On the other hand, from the QCD sum rules one
expects \cite{braun} $\langle x^2\rangle \sim 0.5$ to 1. Therefore, if
the
 AC$^2$M$^2$ {\em ansatz} has chances to survive the
value of $\rho$ must be small. In other words,
$m_{sp}^2 \ll p_F^2$, the spectator must be relativistic. In this
case $F_{Rom}$ is rather broad -- its width is of order $\La$ --
and very asymmetric, see Fig. 10. If $\rho$ is indeed
small, $p_F \approx (\sqrt{\pi}/2)\La$.

It may well happen that the Roman {\em ansatz}, being qualitatively
reasonable in the $b\ra u$ transition, will fail to quantitatively
reproduce fine structure. With  two free parameters it may turn out
to be  too restrictive.  Then we will have to invent another
clever {\em ansatz}, with three fit parameters.

I hasten to add that the AC$^2$M$^2$ model does not reproduce
correctly the $1/m_Q^2$ and higher effects, and it should not --
it was not engineered for that purpose. What may be even more
important, it gives no hints on the
evolution of the light-cone distribution function
towards the temporal one with increasing $m_q$. Therefore,
the description based on this model is generally speaking
inapplicable in the
$b\ra c$ transition; especially detrimental it becomes
in the $B\ra X_ce\nu$ decays at  large
$q^2$, when we approach the SV limit.  As I have mentioned
a few minutes ago a large part of the phase space belongs to the SV
regime. The strongest indication
is the fact that two elastic modes, $B\ra De\nu$
and $B\ra D^* e\nu$, yield more than 60 \% of the total
probability. In the SV regime the dominance of
the elastic modes is absolute.

\section{A few words about hard gluons}

So far non-perturbative effects were my prime concern.
This is the heart of the theory, its non-trivial part.
To make contact with experiment, however, it is necessary
to include a dynamically rather trivial effect due to the hard
gluons, in the bremsstrahlung and in the loops. The bremsstrahlung
corrections are most important in the case of the light final quarks
and are moderate if the final quark is relatively heavy.
Therefore, to make my point, I will consider, as a most instructive
example \cite{bigi4}, the  radiative transition $b\ra s\gamma$
($B\ra
X_s\gamma$ at
the hadron level). As you surely remember, in the parton
approximation
the photon spectrum is a monochromatic line at $E=m_b/2$. The
primordial motion
of the heavy quark, a non-perturbative effect, smears this line and
generates a spectrum in the window and in the adjacent region
below
the window.  The width of the line becomes of order $\La$.
Outside the primordial domain whose size is $\sim \La$ the
primordial
motion has a negligible impact on the photon spectrum.

Even if the primordial motion is switched off, the line will be
smeared anyway due the hard gluon emission. A bremsstrahlung
gluon
can carry away a finite fraction of the accessible momentum,
$\sim m_b/2$. This contribution produces events with the
invariant mass of the hadronic state of order $m_b/2$;
the corresponding photon energy spectrum is ${\cal O}(\alpha_s)$.
The effect is calculable as a one-loop correction in
the transition operator.

Apart from this trivial radiative tail the hard gluon emission
modifies the peak itself, much in the same
way as the electromagnetic radiative corrections modify
the $J/\psi$ peak in $e^+e^-$ annihilation. There the natural
width of the $J/\psi$ meson is negligibly small,
and the observed shape of the peak is  determined
by the radiative smearing due to the photon bremsstrahlung.

The gluon bremsstrahlung is characterized by large logarithms in the
end-point domain; summation of these logarithms results in the
Sudakov exponent. The Sudakov effect destroys the
monochromatic line of the parton model. (It does not populate
the window, though; it clearly remains empty in perturbation
theory).

The peak at $E=m_b/2$ becomes less singular but still a singularity
persists. The
total probability remains the same as in the parton model but, if in
the
parton model it is all saturated in the peak, now a part of it is
pumped into the adjacent tail. The area under the distorted peak
turns out to be $\sim (\La /m_b)^{\epsilon_0}$
where $\epsilon_0$ is a positive but relatively small number,
$\epsilon_0\sim 0.3$ \cite{bigi4}.

For some reasons which I do not understand the problem of the
Sudakov suppression in the end-point domain
is perceived as something intractable:
some authors say that the running coupling constant controlling
these corrections blows up and no reliable calculations are
possible. In principle, these  questions
have been studied in similar problems in the late seventies.
To feel the running nature of $\alpha_s$ it is necessary to go beyond
the double-log level. The answer is well-known at least at the first
subleading level (with one logarithm lost): the coupling
constant entering the Sudakov exponent is the running coupling
normalized at $k^2$ where $k^2$ is the invariant mass squared of
the hadronic state in the corresponding jet. In our case of
the $b\ra s\gamma$ transition $k^2 = (m_bv-q)^2$.

If so, the transformation of the parton-model delta
function is easy to find. The Sudakov logarithms
generate the following spectrum at the end-point:
\beq
d\Gamma /dE \propto \frac{1}{[1-(2E/m_b)]^{1-\epsilon_0}}
\theta (\frac{m_b}{2} - E)
\eeq
where $\epsilon_0 = 8/(3b)\sim 0.3$ ($b$ is the first coefficient
in the Gell-Mann-Low function). This expression neglects subleading
logarithms whose
effect reduces to an overall factor $1+{\cal O}(\epsilon_0)$.
Thus, there is an uncertainty of order ${\cal O}(\epsilon_0)$ in
the normalization of the end-point structure which must
(and can) be eliminated by further calculations.

The final predictions for the spectral distributions, ready for
comparison with experimental data, are obtained
by superimposing the perturbative corrections with the
non-perturbative effects I have discussed previously. The primordial
shape function is smeared, of course, (in a well-defined way):
its basic feature -- a peak in the end-point domain -- stays
intact.

\section{What is the heavy quark mass?}

So far I have deliberately avoided any explicit definition of the
heavy
quark mass $m_Q$, as well as the related parameter $\La$, although
both are the key parameters of the theory, appearing in virtually
every expression. If you start thinking on this issue
you will realize that the question is not that simple as it might seem
at first sight and as it sometimes presented in the literature.
The distinction between different definitions of mass is obviously
a pre-asymptotic effect since the difference is
of order $\Lam$, a $\Lam /m_Q$ effect. Since we are aimed,
however, at a
systematic
theory of pre-asymptotic effects ensuring accuracy
$1/m_Q^2$ and higher we have to address the question of mass
from scientific positions \cite{bigi6}.

Isolated quarks do not exist in nature; therefore, the notion
of an on-shell quark mass is meaningless. Heavy quarks
are always surrounded by a light cloud. How much of it can one peel
off?

The quantity most commonly used to parametrize $m_Q$ is the
so-called pole
mass $m_Q^{pole}$. In perturbation theory the quark Green function
is well
defined in each (finite) order; the pole mass is, by definition, the
position
of the pole of the quark propagator to a given (finite) order in
$\alpha_s$.
Thus, what people actually do is equivalent to separating out the
non-perturbative component of the cloud from its
perturbative (Coulomb)
part.  The perturbative component is supposed to be included in
$m_Q$ while the non-perturbative one to $\La$.

In purely perturbative calculations such a definition is quite
acceptable,
as well as many others. Moreover, it is even more convenient than
others since the gauge invariance of $m_Q^{pole}$ is explicit. This
feature
made it very popular since the mid-seventies (see e.g. Ref. 36).
Important  results
of the perturbative calculations, such as the total semileptonic
widths, are
routinely expressed in terms of the pole mass.

What is acceptable in PQCD becomes absolutely unacceptable
in the theory of non-perturbative effects. As a matter of fact, it is
impossible to define  $m_Q^{pole}$ to the accuracy better than
$\Lam /m_Q$ \cite{bigi6}. The discrimination ``perturbative versus
non-perturbative"
contradicts the spirit and the letter of the Wilsonian
operator product expansion and can not be carried out
in a systematic way. The right thing to do is to separate
out the low-frequency component of the cloud
($\omega <\mu$)
from the high-frequency one ($\omega >\mu$), and include
the high-frequency component in the definition of the quark mass.
Then the low-frequency component will reside in $\La$.
In other words, any scientific approach must deal with
$m_Q(\mu )$, the running mass introduced at a distance
$\mu$ below the would-be pole. As usual, the choice of
the parameter $\mu$ is dictated by two opposite requirements:
on one hand, to  come as close to the soft non-perturbative domain
as possible we want $\mu$
to be as small as possible; on the other hand, to develop a consistent
theory
we must ensure that $\alpha_s (\mu )/\pi \ll 1$. The fact that both
requirements
can be met simultaneously is a highly non-trivial feature of QCD,
established so far only at an empirical level. Practically
$\mu=$ several units $\times\Lam$.

The fact that it is impossible to define $m_Q^{pole}$ to the
accuracy better than $\Lam /m_Q$ is seen in many different ways.
One of the simplest arguments comes from
the consideration of the divergence of perturbation theory
in high orders due to the infrared renormalons [37-39]. Of course,
nobody
claims that the presence of the infrared renormalons is the only
obstruction
to introducing $m_Q^{pole}$, the  {\em only uncontrollable}
non-perturbative piece. But their occurrence is
quite sufficient to kill any consistent calculation
dealing with
$m_Q^{pole}$ and aimed at accuracy $\Lam /m_Q$ or better. The
renormalon
contribution gives an idea of the theoretical uncertainty
which can not be eliminated, in principle, without introducing
$m_Q(\mu )$.

Let us examine the perturbative series for
the heavy quark mass. The renormalon divergence originates
from the ``bubble" insertions into, say, the one loop graph
(Fig. 11). A standard expression for the one-loop renormalization
can be written as
$$
m_Q^{pole}-m_Q(\mu ) =\frac{8\pi\alpha_s}{3}
\int_{|\vec k |<\mu}
\frac{d^3k}{(2\pi )^3}\frac{1}{{\vec k}^2} .
$$
where $\mu$ is the normalization point of the running
mass.
  Summing up all the bubbles we effectively
substitute the bare coupling constant by the running
one in the {\em integrand},
\beq
m_Q^{pole}-m_Q(\mu ) =\frac{8\pi}{3}
\int_{|\vec k |<\mu}
\frac{d^3k}{(2\pi )^3}\frac{\alpha_s ({\vec k}^2)}{{\vec k}^2} .
\eeq
Now, the running gauge coupling is given by
$$
\alpha_s ({\vec k}^2) =
\frac{\alpha_s (\mu^2)}{1-(b\alpha_s(\mu^2)/4\pi)\ln (\mu^2
/{\vec k}^2)} =
\alpha_s (\mu^2 )\sum_{n=0}^{\infty}
\left( \frac{b\alpha_s (\mu^2 )}{4\pi}\ln\frac{\mu^2}{{\vec k}^2}
\right)^n
$$
where $b$ is the first coefficient in the Gell-Mann-Low function.
Substituting this expansion in eq. (38) and doing the $k$ integration
we
arrive
at a factorially divergent series
\beq
m_Q^{pole}-m_Q(\mu )=\frac{4\alpha_s (\mu )}{3\pi}\mu
\sum_n n!\left( \frac{b\alpha_s(\mu )}{2\pi}\right)^n .
\eeq
This series is not Borel-summable; truncating it at an optimal value
of $n$,
$n_0 \sim 2\pi/(b\alpha_s (\mu))$, we find that the intrinsic
uncertainty in $m_Q^{pole}-m_Q(\mu )$ is
\beq
\Delta (m_Q^{pole}-m_Q(\mu ))
\sim \frac{8}{3b}\mu\exp \{ -\frac{2\pi}{\alpha_s (\mu )}\}
\sim \frac{8}{3b}\Lam .
\eeq
(For a related discussion of the infrared renormalons in connection
with HQET, with similar conclusions, see Ref. 40).

Eq. (38) counts the energy of the would-be Coulomb tail around
a static color charge. The tail actually does not exist in the
non-abelian theory; this is the reason explaining the intrinsic
uncertainty (40).

The fact that there
is no scientific way to define $m_Q^{pole}$ to better accuracy
does not mean, of course,
that one can not construct a systematic expansion in $1/m_Q$. This
expansion,
however, must be based on HQET/OPE and explicitly include the
normalization
point $\mu$. Then the expansion parameter is actually $m_Q (\mu
)$.

Although $m_Q (\mu )$ is not directly measurable it is related,
nevertheless,
to observable quantities. Let us elucidate the last point
in more detail.

Assume that the structure functions in the
semileptonic decay  $B\ra X_cl\nu$ are measured separately
as functions of $q_0$ and $|\vec q|$;
we are specifically interested in $w_1$. (I will not bother
you with precise definitions, $w_1$ is just one of five
possible structure functions, see e.g. Ref. \cite{blok2}).  Assume
furthermore
that the measurement can be made at a fixed and small value
of $|\vec q|$, i.e. $|\vec q|\ll M_D$.  Then  the plot of
$w_1$ will schematically look as depicted on  Fig. 12. Qualitatively it
is similar to the plot of Fig. 8 where $E$ is now replaced by $q_0$; a
long radiative tail to the left of the shoulder  was
omitted on Fig. 8. This tail has nothing to do
with the primordial motion, it is due to the hard gluon emission. The
height of elastic delta function is 1, in certain units; a few
conspicuous excitations in the shoulder have heights ${\cal O}(v^2)$
where
$
v=|\vec q|/M_D
$
 while the height of the radiative tail is further suppressed by
$\alpha_s (q_{0max} -q_0)/\pi$.

Using our approach we can formally show \cite{bigi7} that the
integral
\beq
\frac{1}{2\pi}\int (q_{0max} -q_0) dq_0 w_1 = \frac{{\vec
q}^2}{2M_D^2}\, \La
\eeq
modulo corrections of higher order in $v$ and in $\Lam$.
The upper limit of integration is $q_{0max}$, the lower limit
is formally zero.

In these terms the standard definition of a ``universal constant"
$\La$, and the corresponding pole mass, $m_b^{pole} =M_B-\La$,
might be formulated as follows: (i) take the radiative perturbative
tail to the left of the shoulder and extrapolate it all the way to
the point $q_0=q_{0max}$; (ii) subtract the result from the measured
structure function $w_1$; (iii) integrate the difference over $dq_0$
with the weight function $(q_{0max} -q_0)/2\pi$. The elastic peak
drops out and the remaining integral is equal to $(v^2/2)\La$.

When I rephrase the standard program that way its absurdity,
from  purely theoretical point of view, becomes evident.
How can one define what the perturbative tail is at
$(q_{0max} -q_0) \sim \Lam $ even in the first order in $\alpha_s$,
to say nothing about  high orders? Meanwhile, this
is exactly what is implied in the pole mass: one includes
in the pole mass, by definition, all perturbative corrections coming
from all virtual momenta, up to the would-be mass shell (i.e.
excludes them from $\La$).

What is the correct procedure? We {\em must} introduce a
normalization point
$\mu$ in such a way that all frequencies smaller than
$\mu$ can be considered as ``soft" or
inherent to the bound state wave function; at the same time
$\alpha_s(\mu )$ is still a well-defined notion and the perturbative
expansion in $\alpha_s(\mu )/\pi$ makes sense. We then draw a line
at $q_0 =q_{0max}-\mu$ (Fig. 12). The integral (41) taken
in the limits from $q_{0max}-\mu$ to $q_{0max}$ represents
$(v^2/2)\La (\mu)$
modulo corrections of higher order in $v$ and in $\Lam$.
The running mass is then defined as $m_b(\mu )
=M_B - \La (\mu )$. Practically, if the radiative tail
in $w_1$ is numerically suppressed compared to the
height of the shoulder, the $\mu$ dependence of $\La$
may be rather weak.

\section{The onset of duality. Divergence of OPE}

In this part of my talk I will leave  particular  applications
and address a  general issue, the  foundation of the
theoretical approach developed.
The theory of preasymptotic effects in weak decays of heavy flavors
is based on the Wilsonian operator product expansion. More exactly,
we exploit a generalization of OPE relevant to the  Minkowski
kinematics. To
justify the procedure we keep in mind a possibility of analytic
continuation to the euclidean domain in some momenta. In the
semileptonic decays we can analytically continue in the lepton pair
momentum and consider the amplitudes of interest off all cuts, as it
is
done, for instance, in Ref. 10. In the general case it is always
possible to introduce an auxiliary (complex) momentum flowing in
(or out) the vertex generating  the weak decay.  The
operator product expansion is unambiguously defined in the
euclidean domain since here the separation of the large and short
distance contributions is well-defined.  Predictions in the Minkowski
domain are then
obtained through dispersion relations. The simplest prediction of this
type is the calculation of the total $e^+e^-$ annihilation cross section
at high energies.

The coefficient functions in the original (euclidean) version of OPE
are determined by the short distance dynamics. From the remark
above it must  be clear that the very same statement is valid for the
physical observables in the Minkowski region provided
they are fully inclusive and the predictions are
understood in an integral sense,
smeared over some energy range.  A point-by-point prediction
becomes
possible only at sufficiently high energies, when the
observables under consideration
are already smooth and additional ``by hand" smearing is not needed.
The theory of the preasymptotic effects I have discussed
above assumes that
we are already inside this {\em duality domain}. The question of
where
the boundary of the duality domain lies in each particular process
is one of the major practical questions in the non-perturbative QCD.

To make the point more transparent let us turn to the $e^+e^-$
annihilation cross section where the kinematical conditions are much
simpler than in the heavy flavor decays. If $q$ is the total
momentum
of the pair, $q^2>0$ corresponds to the physical (Minkowski) region
while negative values of $q^2$ (i.e. positive $Q^2\equiv -q^2$) is the
euclidean domain.  The operator product expansion in the euclidean
domain is the expansion of the
current correlation function in the powers of $\alpha_s (Q)$  and
$\Lam^2/Q^2$ modulo
logarithms of the same parameter. As well-known
\cite{renormalon,renormalonr} the perturbative
series
is factorially divergent and non-Borel-summable. The standard
perturbation theory has to be amended.

It was amended \cite{SVZ} in the spirit of OPE -- by means of
introduction of the vacuum condensates,
\beq
\langle vac |{\cal O}_n|vac\rangle\sim \mu^{d_n}
\exp\left[ -\frac{2\pi d_n}{b\alpha_s(\mu )}\right]
\times\,\, logs \, ,
\eeq
where $d_n$ stands for the dimension of the operator ${\cal O}_n$.
This rather straightforward extension cures the divergence
of the perturbative series; the exponential (in $1/\alpha_s$)
terms appear explicitly. They are obviously non-analytic in
$\alpha_s$. These terms result in a complicated analytic
structure in the complex $\alpha_s$ plane, with an infinite set
of cuts \cite{renormalon}.

The expansion for the current correlation function at $Q^2>0$
is translated in the prediction for the total cross
section, $\sigma (s)$, where $s=q^2$. In the limit $s\ra\infty$
this prediction is simple:  an
appropriately normalized ratio $R ={\rm Const}\, (s\sigma )$
is equal to unity (I assume for simplicity that the only quarks are the
massless $u$ and $d$).  Of course, we are interested in deviations
from this unity.

Every given term in the euclidean expansion results in a correction in
$R(s)$ of a generic form
\beq
\sum a_n (\Lam^2 /s)^{d_n/2}\times\,\, logs\, .
\eeq
(Pure powers of $1/Q^2$ yield delta function of $s$ and its
derivatives invisible at $s\gg\Lam^2$.) We have calculated a variety
of such terms in the case of the heavy flavor decays where the role
of $s$ is played by $m_Q^2$. It is obvious that at $s\gg\Lam^2$
to any finite order in eq. (43) the prediction for $R(s)$ is a smooth
function of $s$.

The question is where the structures in $R(s)$ come from. The fact
that the structures are there is clearly seen, say, in the multicolor
QCD \cite{Nc}, with $N_c\ra\infty$.  In this limit $R(s)$ actually
represents a comb of infinitely narrow peaks:
the lowest one is due to the $\rho$ meson and the rest are due to its
radial excitations. Even if the number of colors is three the first
excitation shows up in $R(s)$ as a noticeable structure although
numerically $m_{\rho '}^2/\Lam^2 \gg 1$.

The answer to the question above is in the behavior of the power
expansion (43) in high orders \cite{shifman4} (or, better to say,
in the behavior of the analogous euclidean expansion). The power
series {\em per se} turns out to be asymptotic, it has factorially
divergent coefficients. The reflection of this asymptotic nature of the
expansion is the occurrence of the exponential terms
of the type
$$
\exp (-CQ^2/\Lam^2 )
$$
in the euclidean domain. Their analytic continuation to the
Minkowski region provides a structure in $R(s)$. They also
set the rate of approach to the duality domain and determine the
accuracy of the duality relations.

The factorial divergence of the power series may be called
the divergence of the second generation. Accumulation of the
't Hooft singularities (43) produces a more sophisticated singularity
\cite{shifman4}
\beq
\exp\left[ -C{\rm e}^{(4\pi b^{-1}/\alpha_s (Q))}\right] .
\eeq
Unlike the 't Hooft singularities which, in principle, can be revealed
by analyzing multi-loop Feynman graphs, the second-generation
singularity (44) obviously remains undetectable diagrammatically
(in the classical understanding of diagrammar).

The assertions above can be demonstrated in a few different ways.
The
simplest proof I know is as follows. Let us consider  heavy-to-light
quark currents
$$
J_S=\bar Q q,\,\,\, J_P=\bar Q\gamma_5 q\, ;
$$
\beq
J_1 = \frac{1}{2}(J_S+J_P)=\bar Q\frac{1}{2}(1+\gamma_5 )q,\,\,\,
J_2 = \frac{1}{2}(J_S^\dagger-J_P^\dagger)=
\bar q\frac{1}{2}(1+\gamma_5 )Q \, .
\eeq
Later on we will assume that $m_Q\ra\infty$ and $m_q\ra 0$.
Note that $J_2$ is {\em not} hermitean conjugate to
$J_1$; rather  $J_2$ is a chiral partner to $J_1^\dagger$. Therefore,
the correlation function
\beq
\Pi = i\int {\rm e}^{ikx} dx \langle vac |T\{J_1(x), J_2(0)\}|vac\rangle
\eeq
vanishes in perturbation theory if $m_q$ is put  to zero. Thus, there
is no uninteresting background, a feature which should be welcome,
of course.

Formally we can write the correlation function (46)
as a trace,
\beq
\Pi (k^2) = i{\rm Tr}\, \left\{
\frac{1}{2}(1+\gamma_5)
\frac{m_q}{{\cal P}^2-m_q^2+(i/2)\sigma G}\,\,
\frac{m_Q}{({\cal P}+k)^2-m_Q^2+(i/2)\sigma G}
\right\}
\eeq
where ${\cal P}$ is the momentum operator and the trace operation
is defined with respect to this operator. Simplifying a little bit I will
say  that $\Pi$ is the product of two quark Green's
functions in the background gluon field; the averaging over this field
is implied (Fig. 13$a$). The $(1+\gamma_5)$ projector in the currents
$J_{1,2}$
ensures the ``softness" of ${\cal P}$: only very small eigenvalues
of ${\cal P}^2+(i/2)\sigma G$ contribute since otherwise the explicit
factor $m_q$ in eq. (47) annihilates the result in the limit
$m_q\ra 0$ (compare with Ref. \cite{banks}).

Let us focus on the soft gluon fields discarding the effect of
the hard gluon exchanges which is well-understood. First of all, we
should choose our reference
point, the external momentum $k$, in a most advantageous way.  If
$m_Q\ra\infty$ it is clear that the optimal reference point lies
``slightly" below threshold,
$$
k_0 = m_Q -\epsilon,  \,\,\, \vec k =0,
$$
where $\epsilon$ scales with $\Lam$, not $m_Q$.

Taking now the limit $m_Q\ra\infty$, assuming
the ``softness" of ${\cal P}$ (i.e. ${\cal P}\sim \Lam$; this
assumption is clearly not valid for graphs with the hard gluons which
are disregarded anyway) and neglecting all terms suppressed
by powers of $1/m_Q$ we find that
\beq
\Pi (\epsilon ) =-i{\rm Tr}\left\{
\frac{1}{4}(1+\gamma_5)
\frac{m_q}{{\cal P}^2-m_q^2+(i/2)\sigma G}\,\,
\frac{1}{\epsilon -{\cal P}_0}\right\} .
\eeq
The momentum operator ${\cal P}$ is a perfect analog of the
operator $\pi$ from the previous sections. As a matter of fact,
to be fully consistent, I should have replaced it by $\pi$;  and I will
from now on. If $\epsilon\gg\Lam$ one can expand
eq. (48) in $1/\epsilon$. To write the expansion in a compact form it
is
convenient to introduce the vacuum expectation value of
$\bar q q$ where $q$ is the massless quark field. Then, according to
\cite{banks},
\beq
\langle vac|\bar q q|vac\rangle
=\lim_{m_q \ra 0} (-i){\rm Tr}
\left\{
\frac{m_q}{\pi^2-m_q^2+(i/2)\sigma G}
\right\} .
\eeq
Let us agree that in this section the angle brackets will denote
the vacuum averaging. Then
\beq
\Pi (\epsilon )
= \frac{1}{4\epsilon}
\left[
\langle \bar q q \rangle +\frac{1}{\epsilon^2}
\langle \bar q \pi_0^2q \rangle +
\frac{1}{\epsilon^4}
\langle \bar q \pi_0^4 q \rangle + ...\right] .
\eeq
The term with $\gamma_5$ in eq. (48) drops out since the
vacuum state in QCD is $P$-even. What is more important --
a key point in our analysis -- is the disappearance of the
odd powers of $\pi_0$. The vacuum expectation values with
the odd powers of $\pi_0$ all vanish due to the Lorentz invariance.
Another way to see the absence of the odd
powers of $\pi_0$ is to return to eq. (48) and to observe that
it must stay intact under the change of the sign of the momentum
operator, an obvious property of the trace operation.  Certainly, one
could have written the operator product expansion (50) directly
by cutting the light-quark line on Fig. 13$a$ (see Fig.  13$b$).

A question that immediately comes to one's mind is how eq. (50) can
possibly be correct. Indeed, let us assume that the expansion
(50) is convergent. Then it defines a function which is odd under
the reflection $\epsilon\ra -\epsilon$. By no means one can allow the
correlation function $\Pi (\epsilon )$ be odd!

At positive $\epsilon$ we are below the cut, and $ \Pi (\epsilon )$
is analytic in $\epsilon$. At negative $\epsilon$, however, we sit
right on the cut generated by the intermediate physical states
produced by the currents $J_{1,2}$, and  the correlation function
$\Pi (\epsilon )$ develops an imaginary part, a discontinuity across
the cut. Qualitatively we have a pretty good idea of how
${\rm Im}\Pi$ looks like. I will dwell on this point later and now just
state that on physical grounds the behavior of $\Pi (\epsilon )$
at positive and negative values of $\epsilon$ is absolutely different.

What is the solution of this apparent paradox?

Looking at the definition of the function $\Pi$ and the currents (45)
we observe that $\Pi$ is actually a {\em difference} between the
scalar and pseudoscalar channels,
$$
\Pi \propto \langle J_S, J_S^\dagger\rangle
- \langle J_P, J_P^\dagger\rangle .
$$
Therefore, at first sight one might try to  say that the spectral
densities in
these channels are identical and cancel each other in the difference
so that $\Pi (\epsilon )$ has no imaginary part whatsoever.
The expansion (50) itself tells us, however, that this conjecture is
wrong.
A brief reflection  leads  one to an oscillating imaginary part
schematically depicted on Fig. 14.  The oscillations are needed in
order
to kill all even terms in $1/\epsilon$ in the expansion of the
dispersion representation,
\beq
\Pi(\epsilon ) =\frac{1}{\pi}\int _{E_0}^\infty
dE\, \frac{\Im\Pi (E)}{\epsilon +E} .
\eeq

The spectral density on Fig. 14  refers to the real QCD, with three
colors (I remind that it is the difference between
the spectral densities in the scalar and pseudoscalar channels
and, therefore, it need not be positive).
In the multicolor QCD, with $N_c\ra\infty$, we would have, instead,
two combs of infinitely narrow peaks sitting, back-to back,
on top of each other.  Let us defer for a while
the discussion of how the spectral density evolves from $N_c=3$
to $N_c=\infty$.

Returning to the expansion (50), the only logical possibility is to say
that the coefficients of the $1/\epsilon$ expansion are actually
factorially divergent in high orders, so that the expansion is
asymptotic. Then a function defined by this expansion (plus the
correct analytical properties, see eq. (51)) can well be essentially
different at positive and negative values of $\epsilon$, although
superficially it produces an impression of an  odd function of
$\epsilon$. (Examples of such functions are very well known in
mathematics, and I will present one shortly.) In other words,
the matrix elements $\langle \bar q(\pi_0)^{2n} q \rangle$ at
large $n$ behave as
\beq
\langle \bar q(\pi_0)^{2n} q \rangle\sim
(-1)^n\langle \bar q q \rangle (\Lam^2)^n C^{2n}(2n)!\,  ,
\eeq
where $C$ is a numerical constant. The value of this constant is
crucial since it determines the onset of the duality regime.

The appearance of the factor $(-1)^n$ can be explained as follows.
The vacuum expectation value (52) written as an euclidean integral
has
the form
\beq
\langle \bar q(\pi_0)^{2n} q \rangle
=-\lim_{m_q\ra 0} \, (-1)^n{\rm Tr}\, \left[ (\pi_0)^{2n}\,\,
\frac{m_q}{\pi^2+m_q^2+(i/2)\sigma G}\right]_E,
\eeq
where the conventions concerning the euclidean
$\gamma$ and $\sigma$ matrices are borrowed from
\cite{novikov2}.  The subscript
$E$ marks the euclidean quantities. The expression in the square
brackets is well-defined and is positive-definite.

Upon reflection it is not difficult to understand that the factorial
growth of the condensates (52) with $n$ is quite natural. Combining
eqs. (50) and (52) on one hand with the dispersion relation (51), on
the other, we conclude that the absolute value of Im$\Pi$ at large
$E$ is exponential in $E$,
\beq
|{\rm Im}\,\Pi |\sim\exp (-E/(C\Lam ))
= \exp\left[ -\frac{1}{C}{\rm e}^{(2\pi b^{-1}/\alpha_s (E))}\right] .
\eeq
The truncated operator product expansion (50) would predict
Im$\Pi = 0$ at  large $E$. Thus, the exponential in eq. (54) gives an
estimate of the accuracy of the duality relations.  The number $C$
does indeed play a key role since the duality sets in at $E\gg C\Lam$.

Speaking in terms of the analytical structure in the $\alpha_s$
plane we deal here with a sophisticated singularity of the type
exponential of the exponential of $1/\alpha_s$!

I have no time to go into further details. Frankly speaking,
quantitatively, not so many details are worked out. Let me make
only
two remarks.

If the number of colors increases from 3 to infinity the spectral
density Im$\Pi$ at large $E$ experiences a dramatic evolution.
A smooth function depicted on Fig. 14 is converted into two combs of
delta functions -- scalar resonances have positive residues while
pseudoscalars appear in Im$\Pi$ with negative residues;
all residues are ${\cal O}(1)$. (Certainly, averaging
over many resonances produces Im$\Pi\approx 0$.) This means that
the terms in the operator product expansion (50) which are
subleading in $1/N_c$ must be more singular in $n$, i.e. their
divergence in high orders of OPE must be stronger. This aspect is not
new, though; the very same situation takes place in the ordinary
perturbation theory \cite{koplik}.

To wind up the topic let me give an example of a function whose
properties are very similar to those we found for the correlation
function (46). If $\psi (z)$ is the logarithmic derivative of the
$\Gamma$ function a schematic model for $\Pi$ is as follows:
\beq
\Pi_{model} =\psi (\epsilon +\frac{1}{4})-
\psi (\epsilon -\frac{1}{4}) -\frac{1}{\epsilon -(1/4)} .
\eeq

{}From the well-known properties of the $\psi$ functions \cite{wang}
we infer that  $\Pi_{model}$ is a sum of simple poles. All poles
are situated at real negative values of $\epsilon$, {\em viz.}
\beq
\epsilon = -\frac{1}{4},\,\,  -\frac{3}{4},\,\, -\frac{5}{4},\,\,
-\frac{7}{4},\,\,  ...
\eeq
and all residues are $\pm 1$. The signs alternate: the first
pole has negative residue, the second positive and so on. In other
words, the analytic properties of $\Pi_{model}$ are perfect --
analytic everywhere except the points (56) corresponding
to infinitely narrow resonances.

On the other hand, at {\em large positive}
$\epsilon$ our model correlation function is expandable
\footnote{As a matter of fact, the asymptotic expansion (57)
is valid everywhere in the complex plane except a small
angle near the negative real semiaxis, see sects. 3.12 and 3.13
in \cite{wang}.}
in $1/\epsilon$ and, moreover, the expansion looks like an odd
function,
\beq
\Pi_{model} = -\frac{1}{2\epsilon}\times \sum_{n=0}^\infty
\frac{a_n}{(\epsilon^2)^n} ,
\eeq
where $a_n$ are calculable coefficients. At large $n$ they are related
to the Bernoulli numbers $B_n$. As well-known, the latter grow as
$B_n\sim (2n)!$ (see \cite{wang}, page 23), and the coefficients
$a_n$ grow in the same manner. Thus we see that
$\Pi_{model}$ is a perfect example, it is exactly what we want
to get in QCD dynamically.

\section{Conclusions}

Kolya Uraltsev  recently reminded me  about a remark with which
Bjorken concluded his
talk at {\em Les Rencontre de la Valle d'Aoste} in 1990
\cite{bjorken1}.
Bj said then  "... within a year or two it is quite possible that the
language used in describing heavy quark decay phenomenology will
shift away from comparison of data with Model A or Model T, and
instead be phrased in a language which deals with the importance of
a correction of Type X or Type Y...". He was slightly inaccurate in two
respects: elements of the QCD-based theory of preasymptotic effects
had existed before 1990, and it took about four years -- not one or
two -- to complete the first stage of the theory. In all the rest his
anticipation came true -- today any serious work related
to the heavy flavor decays has to start from the $1/m_Q$ expansion
and analysis of different corrections.

There is still a  lot of work to be done.  On the practical side, the
$1/m_Q$
expansion has to be built at least up to the level of $1/m_Q^3$
in all problems of interest. We must create a catalog of the
$H_Q$ expectation values of relevant  operators, much in the same
way as the vacuum condensates were classified and estimated within
the QCD sum rule method. The issue of the gluon
radiative corrections has to be exhaustively worked out.

On the theoretical side, the most important question, the heart of the
OPE/HQET theory,
is understanding the quark-hadron duality: when it sets
in, what deviations one may expect in any given process, etc.
These questions are intimately related to the general structure of
QCD at large distances and the color confinement mechanism. I do not
exclude that  attempts to answer these questions
emerged from the needs of the applied QCD will help
to QCD fundamentalists -- those who still hope that an analytic
solution of the (continuos) four-dimensional quantum
chromodynamics is possible.

\vspace{1cm}

{\large\bf Acknowledgments}\hspace{1cm} I would like to thank
my co-authors, I. Bigi, B. Blok, N. Uraltsev, A. Vainshtein and M.
Voloshin, with whom I shared the pleasure of working on the theory
of preasymptotic effects, for endless enthusiastic discussions of all
topics relevant to the subject. Discussions with B. Grinstein, D. Gross,
I. Kogan,  A. Migdal, N. Nekrasov, M. Neubert, L. Randall and M. Wise
are
gratefully  acknowledged.

This work was supported in part by DOE under the grant
number DE-FG02-94ER40823.

\vspace{1cm}

\section*{Figure Captions}

Fig. 1.\\
Different inclusive transitions of the heavy quark
in the quark language. The closed circle denotes
an effective electroweak vertex. Thick solid line denotes  the
decaying heavy quark.

\vspace{0.3cm}

Fig. 2.\\
Nuclear $\beta$ decay.

\vspace{0.3cm}

Fig. 3.\\
Graphic representation of the transition operator in the inclusive
semileptonic decay. The region inside the dashed line is governed
by short-distance physics.

\vspace{0.3cm}

Fig. 4\\
The diagrams determining the coefficients of the operators
$\bar QQ$ and ${\cal O}_G$ in the transition operator of Fig. 3.

\vspace{0.3cm}

Fig. 5.\\
The photon energy spectrum in the $Q\ra q\gamma$
transition (the parton-model approximation). Due to the two-body
kinematics the photon line is monochromatic. The final quark is
assumed to be massless.

\vspace{0.3cm}

Fig. 6.\\
A realistic spectrum in the inclusive hadronic decay
$H_Q\ra X_q\gamma$.  The kinematical boundary is shifted
to the right of the parton-model line by $\La /2$. The final
quark $q$ is assumed to be massless.

\vspace{0.3cm}

Fig. 7.\\
Graphic representation of the transition operator in the problem of
the end-point spectrum (the Born approximation). All quark lines are
in the background (soft) gluon field.

\vspace{0.3cm}

Fig. 8.\\
Evolution of the spectrum of Fig. 6 as the mass of the final
quark increases (the schematic plot refers to $m_q\sim m_Q/2$).
The effects of the hard gluon bremsstrahlung are not included.

\vspace{0.3cm}

Fig. 9.\\
The photon spectrum in the SV limit ( $m_Q-m_q\ll
m_{Q,q})$. The effects of the hard gluon bremsstrahlung are not
included.

\vspace{0.3cm}

Fig. 10.\\
The Roman model for the distribution function relevant to  $b\ra u$.
The Roman {\em ansatz} has two fit parameters,
the Fermi momentum $p_F$ and the spectator mass $m_{sp}$.
The parameter $\rho$ is defined as $\rho = (m_{sp}/p_F)^2$.

\vspace{0.3cm}

Fig. 11.\\
Infrared renormalon in the pole mass.

\vspace{0.3cm}

Fig. 12.\\
A schematic plot of the structure function $w_1$ in the inclusive
semileptonic decay $B\ra X_c l\nu$ in the limit
$|{\vec q}|/M_D\ll 1$. A long perturbative tail due to the
hard gluon shake-off is depicted to the left of the primordial domain.
The dashed line shows an extrapolation of this tail in the region
where $q_{0max}-q_0\sim \La $.

\vspace{0.3cm}

Fig. 13.\\
Diagrammatic representation for the correlation function
(46). (a) The product of the quark Green functions in
the soft gluon field; (b) Building the operator product expansion:
the light quark line is soft in the limit $m_q\ra 0$ and can be cut.

\vspace{0.3cm}

Fig. 14.\\
A schematic plot of the spectral density for the correlation function
(46).

\end{document}